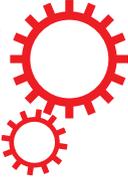

# SCIENTIFIC REPORTS

**OPEN**

# Folding and Stabilization of Native-Sequence-Reversed Proteins

Yuanzhao Zhang[1,2], Jeffrey K Weber[2] & Ruhong Zhou[1,2,3]

Though the problem of sequence-reversed protein folding is largely unexplored, one might speculate that reversed native protein sequences should be significantly more foldable than purely random heteropolymer sequences. In this article, we investigate how the reverse-sequences of native proteins might fold by examining a series of small proteins of increasing structural complexity ($\alpha$-helix, $\beta$-hairpin, $\alpha$-helix bundle, and $\alpha/\beta$-protein). Employing a tandem protein structure prediction algorithmic and molecular dynamics simulation approach, we find that the ability of reverse sequences to adopt native-like folds is strongly influenced by protein size and the flexibility of the native hydrophobic core. For $\beta$-hairpins with reverse-sequences that fail to fold, we employ a simple mutational strategy for guiding stable hairpin formation that involves the insertion of amino acids into the $\beta$-turn region. This systematic look at reverse sequence duality sheds new light on the problem of protein sequence-structure mapping and may serve to inspire new protein design and protein structure prediction protocols.

The foldability of reversed native protein sequences, to the best of our knowledge, has not been investigated in depth. But reversed sequences are distinguished from random amino acid sequences by a number of features that should enhance their foldability. After all, sequence reversal preserves many crucial characteristics of the native protein sequence, such as polar and nonpolar patterning, secondary structural propensity, and general hydrophobicity[1,2]. Such properties will likely act to bias reversed sequences toward native-like folds. The most drastic change introduced by sequence reversal concerns the inversion of connectivity among chiral residues, which can have a subtle influence on the folded structure[3]. For example, since the side chains of $\alpha$-helical amino acids always point roughly in the N-terminal direction[4], reversing the sequence will change the orientations of these side chains within a native-like structure. This altered local geometry can, in turn, affect attributes like the precise packing of the protein's hydrophobic core.

Hydrophobic collapse is the most important driving force in protein folding[5,6]. Kamtekar *et al.* have shown that designed proteins can be coaxed into a desired fold through the careful patterning of polar and nonpolar amino acids alone[7]. A major factor impacting the ability of reversed sequences to assume native-like folds, therefore, likely regards the degree to which the native hydrophobic core is disrupted upon reversal. Indeed, we here show that minimal insertions within reversed sequences can facilitate efficient hydrophobic packing and restore native-like folds in several model proteins.

The reverse sequence foldability problem is particularly interesting when viewed from a protein design's perspective. Current work in protein design can be roughly divided into two classes: *de novo* protein sequence design, wherein a certain structure is built from scratch[7–11]; and perturbative protein design, whereby existing native proteins are altered to achieve new functionalities and enhanced stabilities[12,13]. Efforts in the former category stretch the capabilities of classical force fields to their limits, while attempts in the latter suffer from a lack of flexibility for producing new structures[8,14]. Kuhlman *et al.* have shown that native protein sequences are close to optimal with respect to their folded structures[15], suggesting a likely advantage to using evolved sequences as scaffolds for designing new protein structures. In this light, sequence reversal (either in the context of piecewise reversal or the inversion of full sequences) offers an alternative type of perturbation that has not been widely applied in native protein redesign protocols.

We find that the foldability of the reversed sequences depends on the size and structural characteristics of associated native proteins; the ability of a reversed sequence to adopt a native-like fold is particularly contingent on the flexibility of its native protein's hydrophobic core. In the following sections, we demonstrate these points by interrogating the reverse foldability of small α-helix, β-hairpin, α-helix bundle, and α/β-proteins.

[1]Institute of Quantitative Biology, Department of Physics, Zhejiang University, Hangzhou 310027, China. [2]Computational Biological Center, IBM Thomas J. Watson Research Center, Yorktown Heights, NY 10598, USA. [3]Department of Chemistry, Columbia University, New York, NY 10027, USA. Correspondence and requests for materials should be addressed to R.Z. (email: ruhongz@us.ibm.com)

## Results and Discussion

In this work, we employ both *de novo* structure prediction algorithms[16–20] and molecular dynamics simulations[21–24] (MD) to explore possible folded states of reversed native protein sequences. Sequences are first fed into the platforms PEPFOLD[19] and QUARK[20] to obtain energetically favorable structures, which then seed 100 ns MD simulations used to guide structural stability assessment and, if necessary, subsequent structural refinement. Details concerning the specific protocols used can be found in the Methods section.

**α-helix.** We begin with three sequences known to favor α-helical structures: those corresponding to $F_s$ –peptide ($A_5(AAARA)_3A$)[25], 3 K(I) (($AAAAK)_3A$)[26], and the first α-helical segment of α3D (GSWAEFKQRLAAI-KTRL)[27]. The first two systems are designed peptides featuring repeating patterns aimed at inducing helicity; the third peptide is a segment taken from a naturally evolved helix bundle. Since each amino acid has a particular α-helical propensity[28], a given sequence's tendency to adopt α-helical structures should be largely conserved upon reversal. Indeed, the structure prediction algorithms identify α-helices as favored structures for both the native and reverse sequences of the three systems studied. Due to a lack of stabilizing hydrophobic interactions, single α-helices often become disordered[25] when solvated in water. In lieu of using MD simulations, we thus apply PyRosetta's scoring function[18,29] to evaluate the relative stabilities of predicted α-helices in native and reverse sequences. PyRosetta yields almost identical stability scores for each pair of sequences examined ($-19.9$ vs $-19.8$, $-27.9$ vs $-27.7$, and $-20.1$ vs $-20.2$ for the native and reverse sequences of Fs-peptide, 3 K(I), and the α3D segment, respectively), implying that the formation of α-helices is largely insensitive to sequence directionality.

**β-hairpins.** β-hairpins pose a stringent test of reverse sequence foldability, since the stability of β-hairpins depends on proper packing within a hydrophobic core. The foldability problem, in this sense, becomes more global and imposes substantial constraints on local residue geometries.

*C-terminus of protein G.* The β-hairpin subunit found in the C-terminus of Protein G (PDB ID: 2GB1; sequence: GEWTYDDATKTFTVTE) is an extensively studied mini-protein stabilized by four hydrophobic residues (W43, Y45, F52 and V54) and seven backbone hydrogen bonds[21,30] in its folded state. Packing among W43, Y45 and F52 across two adjacent strands yields a stable hydrophobic core in the native structure.

Structure prediction algorithms suggest that the reverse sequence also favors a short β-hairpin structure. In MD simulations, however, the predicted hairpin quickly melts away, suggesting that the sequence-reversed protein lacks a well-defined native state. A closer look at the predicted sequence-reversed structure reveals that, unlike in the native C-terminus β-hairpin, the three key hydrophobic residues are pointing outward, discouraging effective packing interactions. This mismatch in side chain directionality is derived from the change in hairpin chirality that accompanies the sequence reversal operation. If the reverse sequence is forced to mimic the native configuration, the reversed sequence directionality within that structure implies an inversion of chirality (i.e. if the native hairpin is left-handed, the superimposition of the reverse sequence on that structure would yield a right-handed hairpin). Certain sequences naturally prefer complementary hairpin chiralities, as corroborated by the native C-terminus hairpin of Protein G studied here. If its reverse sequence were placed in a native-like template, the structure would likely suffer from stress caused by the chirality inversion. Baker and coworkers recently reported a metric that relates local protein structures to tertiary motifs and applied these rules to stabilize designed proteins[31]. Inspired by this work, we achieve a favorable chirality in the reverse sequence hairpin by exploiting the 'β-β rule', which claims that the length of the loop between two β-strands largely determines the chirality of the hairpin. Hairpins with short loops (<4 amino acids) strongly prefer left-handed turns, while those with long loops (>4 amino acids) typically come out right-handed. See Fig.1 in[31] for the precise definition of hairpin chirality.

In the current case, the native β-hairpin is connected by a short loop (length = 3) and is left-handed. We thus inserted two amino acids, Asp9 and Ser11 (as determined to be optimal by PyRosetta; see the Methods section), within the loop region to yield the modified sequence ETVTFTKTDASDDYTWEG. The new hairpin adopts a right-handed hairpin, thus relieving the stress caused by the initial inversion of chirality within native-like, sequence-reversed structures. After these surgical sequence modifications, the hydrophobic packing among central residues improves dramatically, and MD simulations suggest that the resulting hairpin structure is very stable (Fig. 1). One insight that can be drawn from this analysis is that reverse sequences will likely retain strong preferences for native tertiary motifs; we can thus design mutational schemes with native structures in mind, allowing us to remedy factors like chirality inversions that interfere with key interactions.

*Trpzip2.* Trpzip2 (PDB ID: 1LE1; sequence: SWTWENGKWTWK) is a short and exceptionally stable β-hairpin designed by Cochran and colleagues[32]. The protein's superb thermal stability is derived from packing between two pairs of tryptophan residues that bridge the native β-strand pair.

Sequence-reversed Trpzip2 suffers from the same chirality-induced strain seen at the C-terminus of Protein G, a condition again ameliorated by the insertion of two amino acids in the loop region (Fig. S1). The modified sequence adopts a markedly flatter conformation than does native Trpzip2, likely because of slightly different packing patterns among tryptophans. In particular, tryptophan packing within the native protein more closely resembles an alternating, zipper-like arrangement that, requires the two strands to be tightly twisted. Due to the significant stability enhancement provided by tryptophan-tryptophan packing, successful sequence modification should be quite robust to the specific choice of inserted amino acids. Two alanine residues are sufficient to support significant hairpin population. As shown in Fig. 2, the β-sheet components of the modified reverse sequence remain well-defined throughout our simulation.

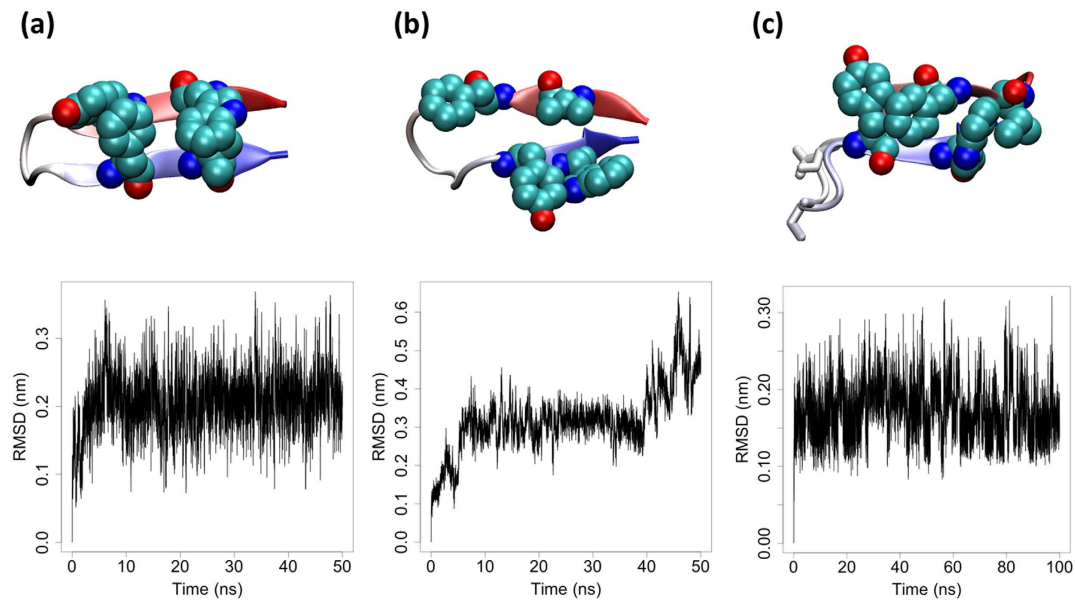

**Figure 1.** Predicted structures and backbone RMSDs for Protein G's (**a**) native β-hairpin (GEWTYDDATKTFTVTE), (**b**) sequence-reversed β-hairpin (ETVTFTKTADDYTWEG), and (**c**) modified sequence-reversed β-hairpin (ETVTFTKTDASDDYTWEG). The residues in bold/red font were inserted into the reverse sequence to improve hydrophobic packing within the hairpin's core. Key residues are rendered as vdW spheres and the residues inserted into the sequence in (**c**) are indicated with white sticks. Hydrophobic residues in the structure predicted for the reverse sequence are not well packed, leading to subsequent deviations in backbone RMSD over the course of MD simulations (**b**). By contrast, the (**a**) native and (**c**) modified hairpin structures are both stable in solution, experiencing only small fluctuations during simulation.

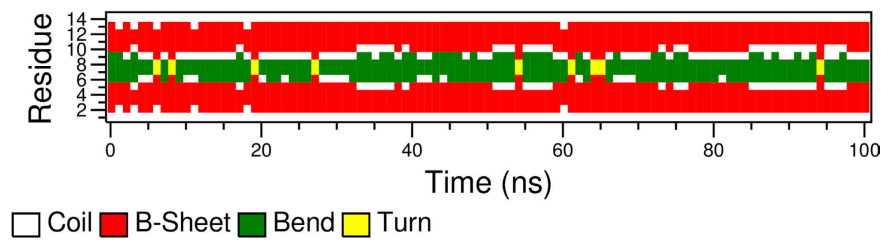

**Figure 2. Secondary structural components of the modified, sequence-reversed Trpzip2 as a function of time.** The hairpin fold remains stable and well-defined for the entire length of the simulation.

**α-helix bundles.** *α3D.* α3D (PDB ID: 2A3D) is a fast-folding three-helix bundle designed to be stabilized by the packing of hydrophobic side chains alone[27]. $S^2$ order parameters for side-chain methyl groups in α3D are quite low[33], indicating that its core is on the fluid end of the packing spectrum and thus does not require very precise hydrophobic interactions to be maintained. This mobility affords native-like structures a certain degree of tolerance to the local geometrical changes introduced by sequence inversion, making α3D a reasonable candidate for stable reverse sequence folding.

Indeed, the reverse sequence favors a native-like, three-helix bundle fold in structure prediction algorithms that remains stable in our MD simulations. Over the course of simulations, the second and third α-helices are quite well-defined; the first segment experiences some moderate fluctuations, but still retains a large α-helical propensity (Fig. S2). To further characterize this sequence-reversed structure, we calculated the solvent accessible surface area (SASA) of the configuration and of its hydrophobic residues (LEU, VAL, ILE, PRO, PHE, MET, TRP), and we compared the results with native α3D (Fig. 3). Most of the hydrophobic residues of the reverse sequence remain buried in the core throughout the simulation, contributing an average SASA of around 6 $\dot{A}^2$ out of a total 55 $\dot{A}^2$. This value compares favorably with the SASA of hydrophobic residues in native α3D, which averages approximately 5 $\dot{A}^2$.

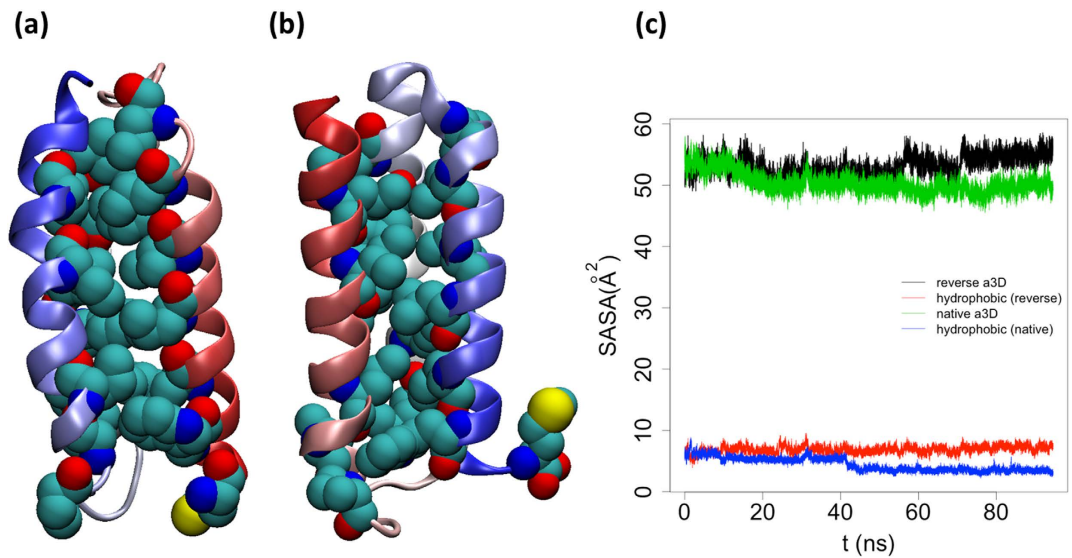

**Figure 3.** (**a**) Structure of native α3D; hydrophobic residues are shown as vdW spheres. (**b**) Structure of sequence-reversed α3D. (**c**) Solvent accessible surface areas (SASAs) of sequence-reversed and native α3D structures and their component hydrophobic residues. Hydrophobic residues within the reverse sequence stay buried throughout the simulation, featuring a SASA comparable to that associated with native α3D (particularly with respect to the crystal structure).

*Trp-cage.* Trp-cage (PDB ID: 1L2Y; sequence: NLYIQWLKDGGPSSGRPPPS) is a fast-folding mini-protein designed by Neidigh *et al.*[34]. Its native structure features an N-terminal α-helix and a C-terminal polyproline II helix connected by a short $3_{10}$-helix motif[22]. Within Trp-cage's native configuration, Trp6 (belonging to the α-helix) is buried between three prolines from the polyproline II helix and another proline from the $3_{10}$-helix to form a tight hydrophobic core. Unlike α3D, Trp-cage is highly optimized and relies on precise packing among Trp6 and Pro12, Pro17, Pro18 and Pro19 to remain stable. It is also much smaller than α3D, leaving less room for structural flexibility upon reversal.

Trp-cage's reverse sequence favors a β-hairpin fold according to QUARK but adopts a native-like fold according to PEPFOLD. The native-like structure predicted by PEPFOLD was determined to be unstable: the proline residues fail to properly bury Trp6, causing the α-helix to dissemble within 10 ns. Pursuing the strategy presented in the β-hairpin section, we attempted to restore the stability of the native structure by inserting amino acids at the intersection of the α-helix and $3_{10}$-helix or by shifting the location of Trp6, hoping to facilitate improved packing between Trp6 and the prolines.

The ten distinct sequence modifications we performed (see Table S1, Fig. S3 and SI text for detailed information) each favored a native-like structure according to both QUARK and PEPFOLD, but none turned out to be stable in MD simulations. Trajectory analysis suggests that the orientational change in Trp6 occurring upon sequence reversal complicates the stabilization of native-like structures. In native Trp-cage, Trp6 points downward, allowing the three C-terminal prolines (Pro17, Pro18, Pro19) to rest above it without creating steric clashes. In sequence-reversed Trp-cages, Trp6 is pointing upward, occupying the space typically occupied by the prolines in the native structure. Based on one representative trajectory (Fig. 4 and Supporting Movie 1), it is evident that the three prolines of interest have to push the upper helix aside to facilitate interactions with Trp6, a need exacerbated by the bulky Tyr3 located one turn above the central tryptophan. The reverse sequence of Trp-cage could thus not be stabilized in a native-like configuration using the sequence modifications tested here.

**α/β protein.** *Protein G.* Protein G (PDB ID: 1MI0) is a 56-residue α/β protein consisting of a four-stranded β-sheet and a single α-helix packed tightly in complex[35]. It has been reported that six hydrophobic residues (Y3, L5, F30, W43, W45, F52) serve as a nucleus during Protein G folding, playing a central role in stabilizing the native structure[36]. Upon reversing the sequence, we find the protein adopts a stable fold very similar to the native structure. The six key hydrophobic residues are not as tightly packed as they are in native Protein G, but other hydrophobic residues (such as ILE7, PHE14, and VAL29) compensate for the deficit, yielding a well-defined hydrophobic core. Key residues for native Protein G and sequence-reversed Protein G are rendered as red sticks in Fig. 5a,b, respectively. The sequence-reversed structure remains intact over the course of our 100 ns MD simulation; its resemblance to the native fold is further demonstrated by the contact maps in Fig. 5. These results for Protein G, considered in concert with those for α3D, suggest that the reverse sequences of larger proteins may be more foldable due to the greater abundance of alternative packing modes that longer sequences facilitate.

**Random native sequence as control.** As a final control experiment, we also evaluated random sequence realizations for proteins which, upon (modified) sequence reversal, adopted stable, native-like folds: α3D, the β-hairpin (C-terminus of Protein G), and Protein G. See the Methods section for details concerning the randomization protocol. Unlike reverse native sequences, these random sequence variants all favored unstable or non-native-like structures. Within metastable conformations that did emerge from random sequences, over half of the secondary structures present were either coils or bends — a structural distribution atypical of globular proteins. These results (illustrated in Fig. S4 and Fig. S5) provide further confirmation that better-than-random patterning is conserved under native sequence reversal.

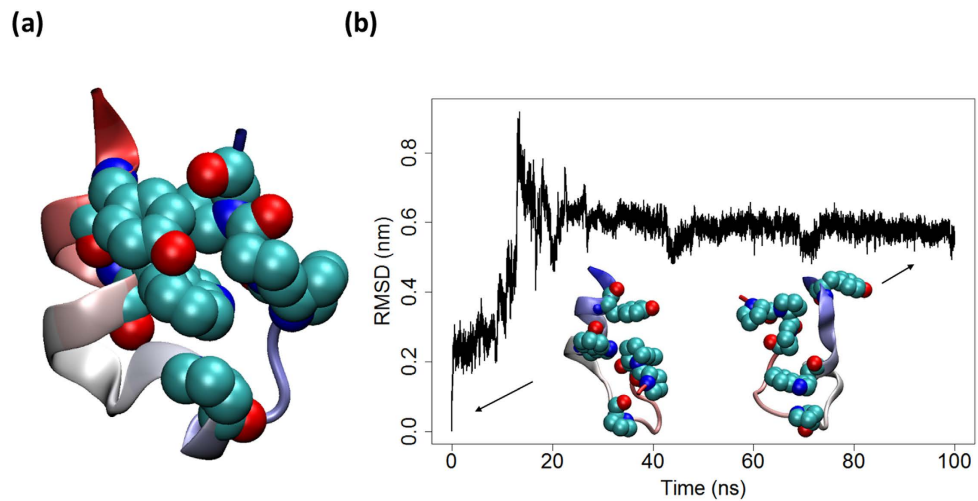

**Figure 4.** (**a**) Native structure of Trp-cage (NLYIQWLKDGGPSSGRPPPS). (**b**) Backbone RMSD of one of the modified, sequence-reversed Trp-cages (SPPPRGSSPGGDKWLQIYLN). The insets provide representative snapshots showing the proline tail deforming the upper helix while trying to bury Trp6. Plotting schemes mirror those used in Fig. 1.

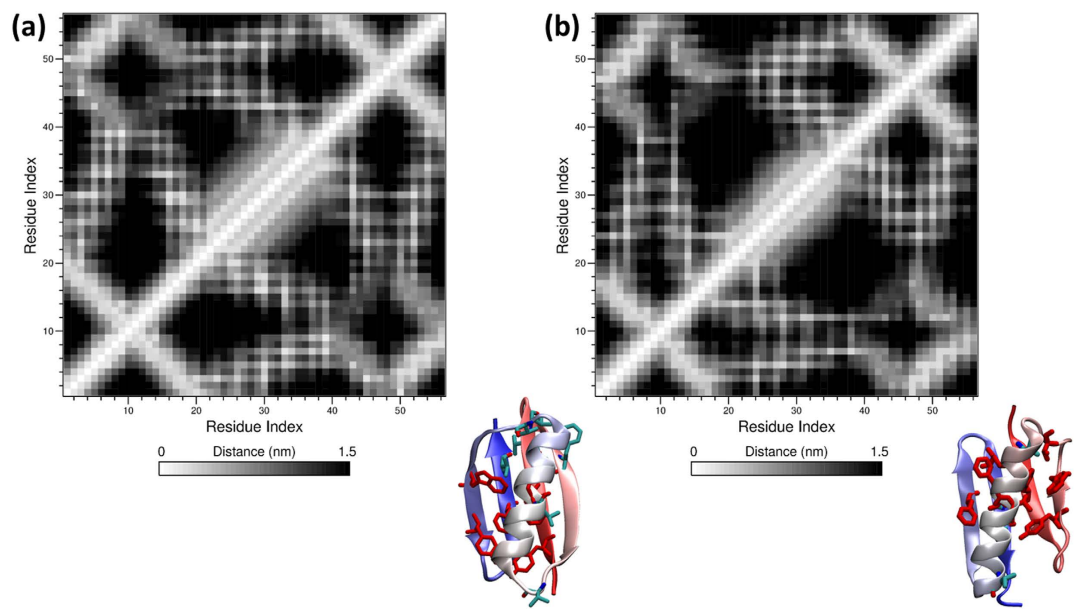

**Figure 5.** Structures and contact maps for (**a**) native Protein G and (**b**) sequence-reversed Protein G. Hydrophobic residues are shown in a stick representation, and residues key to each structure's hydrophobic core are highlighted in red. The mean minimum residue distances presented in the contact maps are calculated by averaging over the whole 100 ns MD trajectory.

## Conclusions
In this work, we have investigated the foldability of reverse protein sequences that feature diverse folds ($\alpha$, $\beta$, $\alpha/\beta$) in associated native states. We found that proteins with relatively mobile hydrophobic cores favor native-like reverse sequence folds. Moderately-sized proteins, such as $\alpha$3D and Protein G, exhibited particularly stable and native-like reverse sequence folds enabled by an array of alternative packing modes emerging within the proteins' hydrophobic cores. For smaller proteins that rely on more precision in core packing, reverse sequences can fail to maintain a well-defined fold due to orientational changes in key hydrophobic residues that occur upon sequence reversal. For simple β-hairpins, we demonstrated the tractability of guiding reverse sequences into native-like structures through the insertion of residues into the turn region.

This study provides a systematic look at the question of reverse sequence folding, which reveals that reverse sequences are perhaps more foldable than intuition alone would suggest. Though our results are encouraging, longer simulations and complementary experiments are needed to place the data described here on a firmer ground. In particular, it remains to be seen whether the (modified) native-sequence-reversed proteins studied in this work retain their structures beyond 100 nanoseconds of simulation time, or whether their features are robust when simulated with other force fields. We also urge further studies to be carried out on the reverse sequences of larger proteins featuring even more complex folds. For larger proteins with hundreds of amino acids, specific polar interactions may play a critical role in defining the structure of native-sequence-reversed proteins, as Huang *et al.*'s recent effort toward the *de novo* design of TIM barrels suggests[37]. Progress in this direction should augment our understanding of protein folding from a unique perspective and provide valuable data for the protein structure prediction and protein design communities. Though simply reversing a given protein's sequence may offer little utility to protein design efforts, fragment-wise sequence reversals, for example, could drive a useful perturbative scheme for integrating stable and diverse structural elements into the modular design of proteins[38,39].

## Methods

**Molecular dynamics simulation.** All MD simulations were performed using the GROMACS simulation package[40] with the OPLS-AA force field[41] at 300 K. Native protein structures were obtained from the Protein Data Bank and solvated with TIP3P water molecules[42] extending at least 1 nm from any protein atom. The systems were equilibrated for 400 ps under the NVT and NPT ensembles before production runs were performed using Berendsen temperature and pressure coupling[43]. The time step for production runs was set at 2 fs with the LINCS algorithm[44] and particle-mesh Ewald method (PME)[45] engaged. van der Waals interactions were treated with a smooth cutoff distance of 1 nm.

**PyRosetta.** For α-helixes, we first applied the standard Rosetta refinement protocol[16] to relax the predicted structures and score them using the Rosetta energy functional with a default weight set for each energy term[29,46].

For the C-terminus of Protein G, we optimized the inserted amino acids by using the design packer task in PyRosetta (restricted within the inserted region), which is based on the Monte Carlo optimization routine in Rosetta and tests side chain packing across a rotamer set that includes all amino acid types. The resulting structure of the optimized sequence was used to seed a subsequent MD simulation.

**QUARK and PEPFOLD.** QUARK is a popular *de novo* protein structure prediction algorithm based on replica-exchange Monte Carlo simulations and an atomic-resolution, knowledge-based force field[20]. It can be accessed at http://zhanglab.ccmb.med.umich.edu/QUARK/. PEPFOLD is another *ab initio* protein structure prediction sever located at http://bioserv.rpbs.univ-paris-diderot.fr/services/PEP-FOLD/, which exploits a greedy algorithm and a coarse-grained force field[19].

**Random native sequences.** We generated three independent random variants of α3D, the β-hairpin (C-terminus of Protein G), and Protein G by inputting each native sequence into Python's random.shuffle function (https://docs.python.org/2/library/random.html). The resulting randomizations of the native sequences (9 in total) were then investigated using the same *de novo* protein structure prediction algorithms and MD simulation protocols employed while evaluating reverse sequences.

## Acknowledgements

We thank Seung-gu Kang, Binquan Luan, Hongsuk Kang, and Bruce Berne for helpful discussions. YZZ is grateful for all the help received from the entire Zhou group during his summer internship at the IBM Watson Research Center. RZ acknowledges the support from IBM Blue Gene Science Program.

## Author Contributions

R.Z. and Y.Z. conceived and designed the research. Y.Z., J.K.W. and R.Z. co-wrote the manuscript. Y.Z. carried out the molecular dynamics simulations and analyzed the data. All authors discussed the results and commented on the manuscript.

## Additional Information

**Supplementary information** accompanies this paper at http://www.nature.com/srep

**Competing financial interests:** The authors declare no competing financial interests.

**How to cite this article**: Zhang, Y. *et al.* Folding and Stabilization of Native-Sequence-Reversed Proteins. *Sci. Rep.* **6**, 25138; doi: 10.1038/srep25138 (2016).